\documentclass[]{spie}  %>>> use for US letter paper
\usepackage{aas_macros}
\usepackage{amsmath,amsfonts,amssymb}
\usepackage{graphicx}
\usepackage[colorlinks=true, allcolors=blue]{hyperref}

\title{Black Hole Spacetime and Properties of Accretion Flows and Jets Probed by Black Hole Explorer \\
--- Science Cases Proposed by BHEX Japan Team ---}

\author[a]{Tomohisa Kawashima}
\author[b,c]{Yuh Tsunetoe}
\author[c]{Ken Ohsuga}
\author[d,e]{Motoki Kino}
\author[f,g]{Yosuke Mizuno}
\author[h]{Kotaro Moriyama}
\author[j]{Hiromi Saida}
\author[k,l,b]{Kazunori Akiyama}
\author[m]{Kazuhiro Hada}
\author[n]{Kotaro Niinuma}

\affil[a]{Institute for Cosmic Ray Research, The University of Tokyo, 5-1-5 Kashiwanoha, Kashiwa, Chiba 277-8582, Japan}
\affil[b]{Black Hole Initiative, Harvard University, \ 20 Garden street, Cambridge, \ MA 02138, USA}
\affil[c]{Center for Computational Sciences, University of Tsukuba, \ 1-1-1 Tennodai, Tsukuba, \ Ibaraki 305-8577, Japan}
\affil[d]{Kogakuin University of Technology \& Engineering, Academic Support Center, 2665-1 Nakano-machi, Hachioji, Tokyo 192-0015, Japan}
\affil[e]{National Astronomical Observatory of Japan, 2-21-1 Osawa, Mitaka, Tokyo 181-8588, Japan}
\affil[f]{Tsung-Dao Lee Institute, Shanghai Jiao Tong University, 520 Shengrong Road, Shanghai 201210, China}
\affil[g]{School of Physics and Astronomy, Shanghai Jiao Tong University, 800 Dongchuan Road, Shanghai
200240, China}
\affil[h]{Instituto de Astrofísica de Andalucía—CSIC, Glorieta de la Astronomía s/n, E-18008 Granada, Spain}
\affil[j]{Daido University, 10-3 Takiharu-cho, Minami-ku, Nagoya, Aichi 457-8530, Japan}
\affil[k]{MIT Haystack Observatory, Westford, MA 01886, USA}
\affil[l]{Mizusawa VLBI Observatory, National Astronomical Observatory of Japan, Iwate 023-0861, Japan}
\affil[m]{Nagoya City University, Aichi 467-0001, Japan}
\affil[n]{Graduate School of Sciences and Technology for Innovation, Yamaguchi University, Yamaguchi 753-8512, Japan}

\authorinfo{Further author information: (Send correspondence to Tomohisa Kawashima, Yuh Tsunetoe, and Ken Ohsuga)\\
Tomohisa Kawashima: kawshm@icrr.u-tokyo.ac.jp\\  
Yuh Tsunetoe: ytsunetoe@fas.harvard.edu\\
Ken Ohsuga: ohsuga@ccs.tsukuba.ac.jp 
}

\begin{document} 
\maketitle

\begin{abstract}
Black Hole Explorer (BHEX) is a space VLBI mission concept, which can probe the black hole spacetime and the plasma properties including the magnetic fields of the accretion flows and relativistic jets. We propose science cases anticipated to be addressed by BHEX mainly via the imaging of the target objects, whose observational features appear in several microarcsecond scale. An appearance of a crescent-shaped shadow in a bright state of the M87 will be able to constrain the magnitude of the black hole spin. A possible appearance of the plasma injection region in the vicinity of the black hole results in the formation of the multiple ring structure and may enable us to understand the jet formation processes. In addition, The separation of linear and circular polarization fluxes and reversal of circular polarization will constrain the magnetic field structure and the thermal properties of the electrons, respectively. Other topics including the test of the gravitational theory are also being discussed. 
\end{abstract}


\section{INTRODUCTION}
\label{sec:intro}  % \label{} allows reference to this section

With the black hole shadow images resolved by Event Horizon Telescope (EHT)\cite{EHTC_M87_2019a, EHTC_SgrA_2022a, EHTC_M87_2024a}, the time has come for us to study near-horizon scale science with $\sim 10 r_{\rm g}$\footnote{Here, $r_{\rm g} (= GM/c^2)$ is the gravitational radius, $G$ is the gravitational constant, $M$ is the mass of the black hole, and $c$ is the speed of light. }.
Attributed to the EHT observations, important results have been obtained: The mass of the black hole is strongly constrained\cite{EHTC_M87_2019f, EHTC_SgrA_2022d}, strongly magnetized accretion flow model (referred to as magnetically accretion disk\cite{igumenshchev_2003, narayan_2003}) is favored to explain the polarized images of M87\cite{EHTC_M87_2021a,EHTC_M87_2021b} and various observation features including lower frequency band images, multi-wavelength spectra and polarimetory in Sgr A*\cite{EHTC_SgrA_2022e, EHTC_SgrA_2024a,EHTC_SgrA_2024b}, and so on.  

The true horizon scale resolution with $\sim r_{\rm g}$ is, however, required to acquire the strong constraint of the black hole spin and unveiling jet formation mechanism as well as the detailed dynamics of accretion flows.
The Black Hole Explorer (BHEX)\cite{BHEX_Johnson_2024, BHEX_akiyama_2024} is a space VLBI mission concept, which can address this grand challenges mainly via the detection of the photon ring (e.g., Refs.~\citenum{johnson_2020}). 
Here, the photon ring is a ring-like image formed mainly attributed to the effect of the strong gravitational lensing in the vicinity of the black holes (e.g., Refs.~\citenum{bardeen_1973}).

Not only by the direct detection of the photon ring itself, the BHEX is also expected to probe the black hole spacetime and the extremely energetic plasma properties including the magnetic fields of the accretion flows and relativistic jets. We propose science cases to be addressed by the BHEX mainly via the imaging of the target objects, which will be complementary studies to the main missions of the BHEX.

\section{SCIENCE CASES PROPOSED BY BHEX JAPAN TEAM}

\subsection{Crescent-Like Shadow --- Another Way to Constrain the Black Hole Spin ---}
\label{sec:crescent}

\begin{figure} [hb]
 \begin{center}
   % \begin{tabular}{c} %% tabular useful for creating an array of images 
   \includegraphics[height=6.5cm]{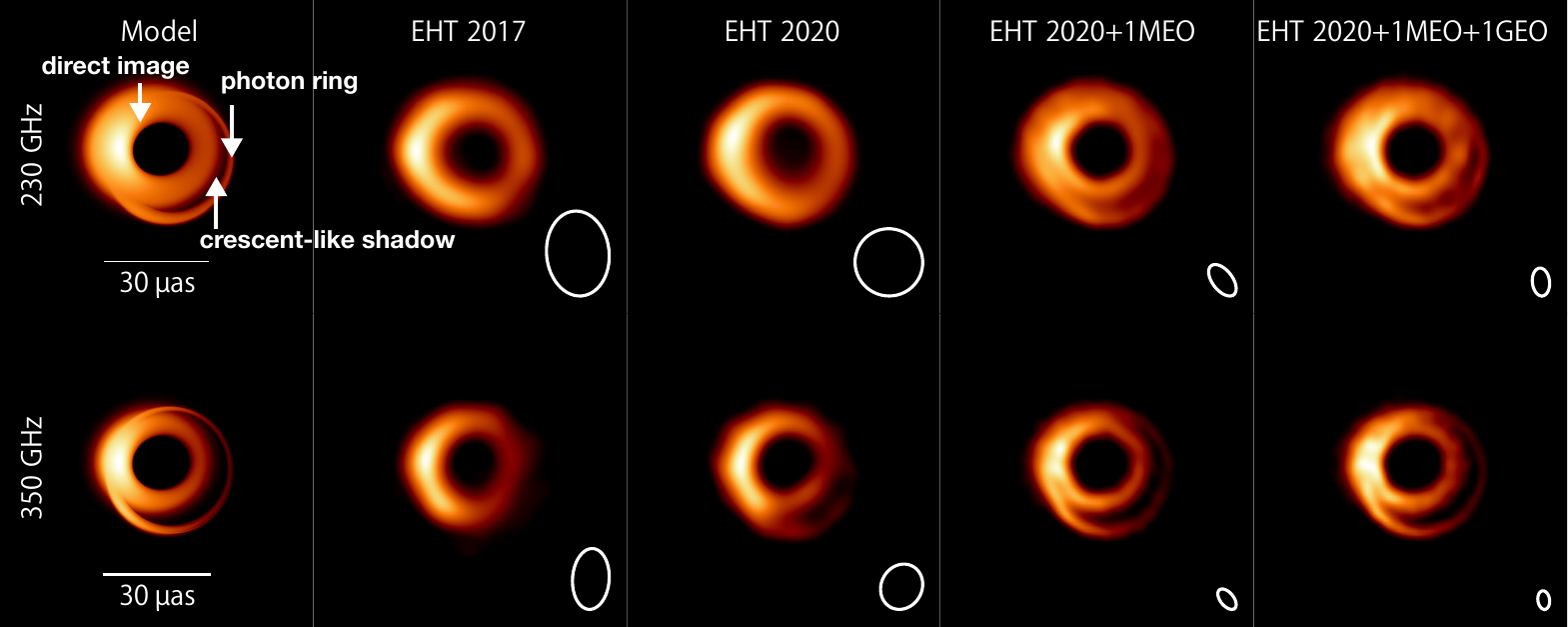}
   % \end{tabular}
 \end{center}
    \caption[example] 
%>>>> use \label inside caption to get Fig. number with \ref{}
   { \label{fig:crescent} 
The theoretical image of crescent-like shadow at 230 GHz (top) and 350 GHz (bottom), and the expected observational images reconstructed by assuming previous EHT array\cite{EHTC_2019} and future ones with space VLBI. (Figure from Refs. \citenum{kawashima_2019}).}
   \end{figure}

The “crescent-like shadow”, as which we refer to the dark region formed between the photon ring and direct image of the accretion flow (Fig. \ref{fig:crescent}), will appear in brighter state ($\sim 1$ Jy) in M87\cite{kawashima_2019}\footnote{In practice, an analysis of early EHT data suggests that the center of M87 may be partially optically thick against synchrotron absorption\cite{kino_2015}. This indicates that the inner part of the direct image of the inner accretion flow is favored to be detected.}.
This will enable us to constrain the black hole spin since the width of the crescent-like shadow depends on the black hole spin. 

% Note: If compiling with LaTeX+dvipdf, please ensure images generated from 
% other software packages have their bounding boxes set correctly.

The dependence of the width of the crescent-like shadow can be explained as follows: The frame-dragging effect (i.e., the effect of the black hole spin) shifts the center of the photon ring, while the center of the direct emission image is less affected by the frame-dragging. As a result, the deviation of the center of the direct image and the photon ring becomes more remarkable as the magnitude of the spin increases.

While the constraint of the black hole spin using the crescent-like shadow is proposed for M87, the same approach will be potentially applied to Sgr A*. The detection of the crescent-like shadow inevitably requires decomposition of the photon ring and the direct image, i.e., the BHEX is needed. This study will be complementary to the main mission of BHEX (i.e., to strongly constrain the spin parameter via the detection of photon ring) \cite{johnson_2020, gralla_2020, palumbo_2023, lupsasca_2024, BHEX_Lupsasca_2024, BHEX_Galison_2024, BHEX_Kawashima_2024}.

\subsection{Multiple-Ring Images of Injected Plasma Jet 
--- Probing Jet-Formation Process ---}
\label{sec:jet_formation}

The formation mechanism of the relativistic jet is still unveiled. One of the most plausible models, the Blandford-Znajek (BZ) process \cite{blandford_1977} requires not only the black hole spin but also the strong magnetic field to explain the estimated jet power in active galactic nuclei. Therefore, the plasma injection process into the highly magnetized jets should be unveiled\cite{blandford_1977,moscibrodzka_2011,broderick_2015,kisaka_2020, kimura_2022}.

Recently, it has been demonstrated that a large ($\sim 60 \mu {\rm as}$ in M87 when the normalized black hole spin parameter is $a_{*} \sim 0.9$) and possibly faint multiple ring-like images will appear if the plasma is injected on and near the separation surfaces connecting inflows and outflows inside the highly magnetized jets\cite{ogihara_2024} (see also Refs. \citenum{kawashima_2021}). The detection of these rings will enable us to constrain the plasma injection scenario and possibly the jet formation mechanism.

% Note: If compiling with LaTeX+dvipdf, please ensure images generated from 
% other software packages have their bounding boxes set correctly.
   \begin{figure} [ht]
 \begin{center}
   % \begin{tabular}{c} %% tabular useful for creating an array of images 
   \includegraphics[height=9cm]{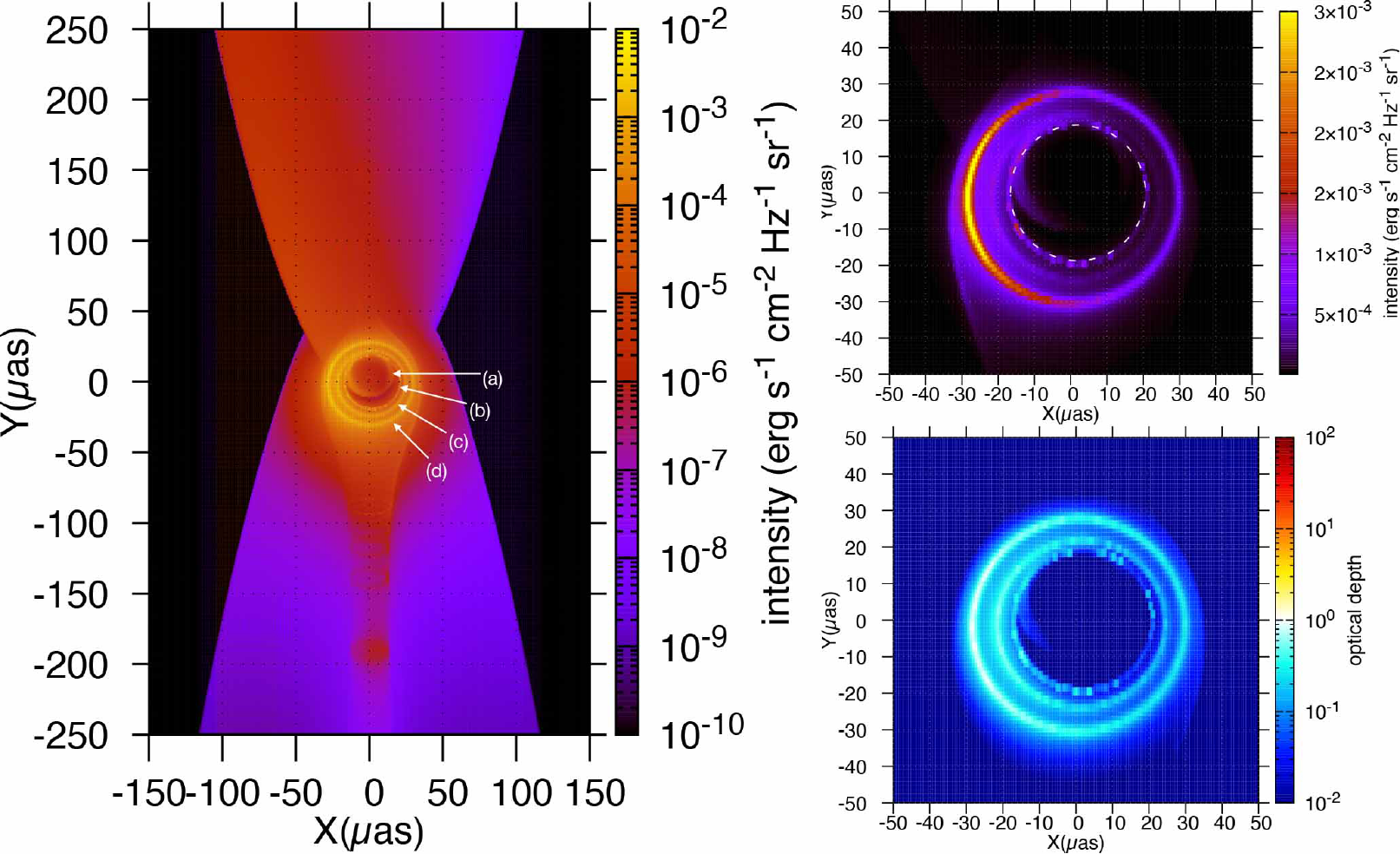}
   % \end{tabular}
 \end{center}
    \caption[example] 
%>>>> use \label inside caption to get Fig. number with \ref{}
   { \label{fig:jet_formation} 
The multiple ring-like images originated from the separation surface of the highly magnetized jets. (Figure from Refs. \citenum{ogihara_2024}).}
   \end{figure} 

\subsection{Polarization Features \\
--- Imprints of Magnetic Field Structures and Electron Heating/Acceleration --- }
\subsubsection{Separations of Linear and Circular Polarization Peak-Fluxes on the Ring Images}
\label{linear_circular}

\begin{figure} [ht]
 \begin{center}
   % \begin{tabular}{c} %% tabular useful for creating an array of images 
   \includegraphics[height=7cm]{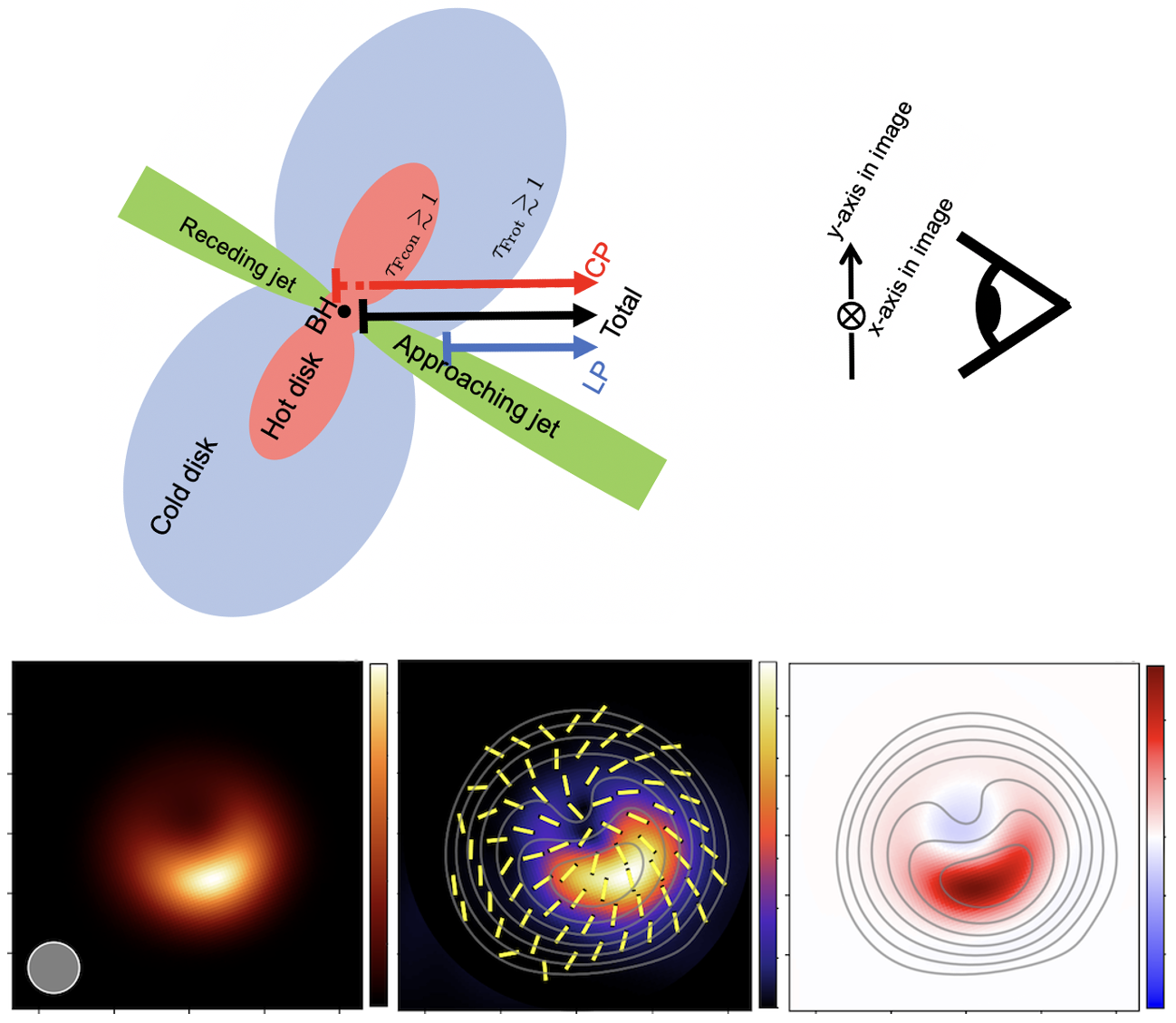}
   % \end{tabular}
 \end{center}
    \caption[example] 
%>>>> use \label inside caption to get Fig. number with \ref{}
   { \label{fig:LPCPsep} 
Top: schematic picture of LP-CP separation. 
Bottom: Theoretical images of total intensity (left), LP (center), CP (right) at 230 GHz. (Figure from Refs.~\citenum{tsunetoe_2022})}
   \end{figure}

Synchrotron polarization is affected by Faraday effect in propagation around a black hole; Linear Polarization (LP) vectors are scrambled by Faraday rotation, and Circular Polarization (CP) components are increased by Faraday conversion. As a result, they show different distributions on the ring, LP-CP separation\cite{tsunetoe_2022,Tsunetoe_2022b} (Fig. \ref{fig:LPCPsep}). Through this feature, we can survey the magnetic fields, electron temperature near black hole, and jet launch mechanism.

% Note: If compiling with LaTeX+dvipdf, please ensure images generated from 
% other software packages have their bounding boxes set correctly.

\subsubsection{Polarization Flips: Imprints of Non-thermal Electrons}
\label{polarization_flip}

\begin{figure} [ht]
 \begin{center}
   % \begin{tabular}{c} %% tabular useful for creating an array of images 
   \includegraphics[height=5cm]{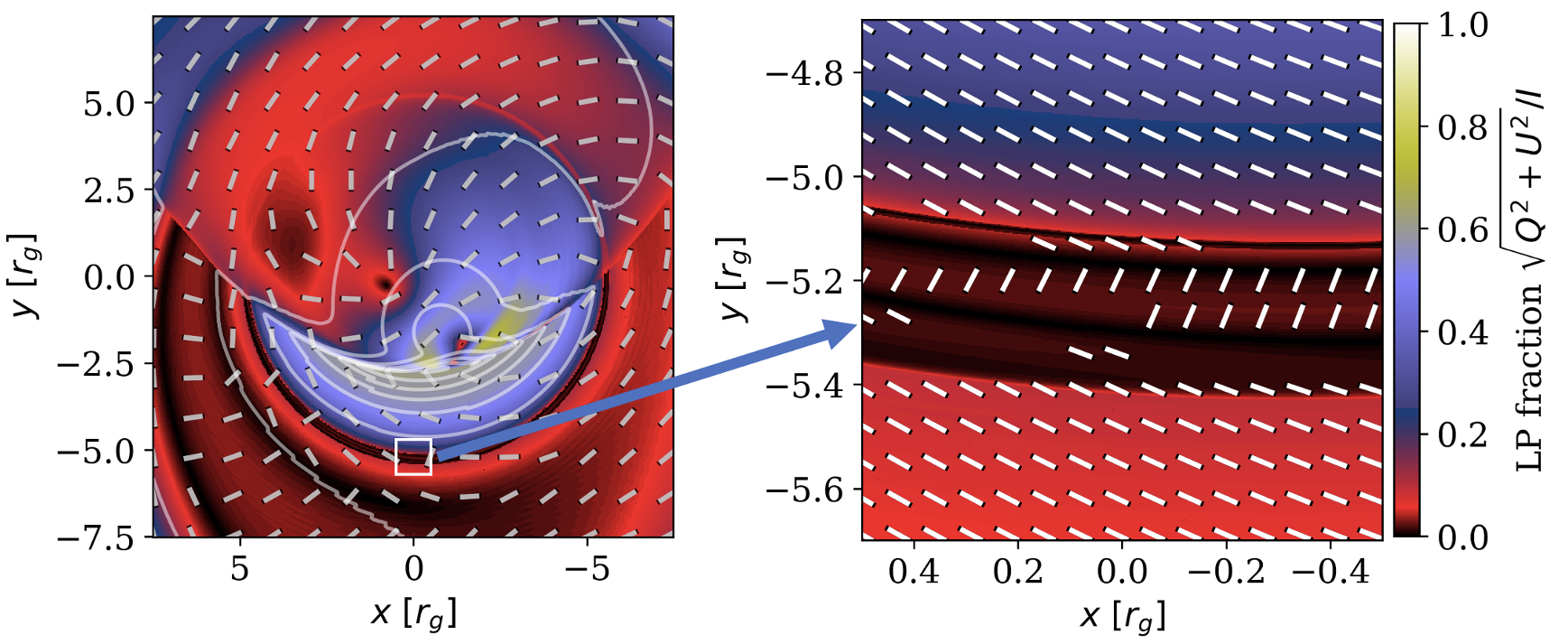}
   % \end{tabular}
 \end{center}
    \caption[example] 
%>>>> use \label inside caption to get Fig. number with \ref{}
   { \label{fig:LPflip} 
Left: Theoretical images of LP at 86 GHz, based on a non-thermal jet model. LP vectors are denoted by the ticks. Right: it's zooming up onto the photon ring (Figure from Refs.~\citenum{tsunetoe_2024})}
   \end{figure}

Particle injection and acceleration around a black hole can lead to non-thermal energy distribution. It is known that opaque non-thermal electrons produce 90°-flipped LP vectors and oppositely-handed CPs, compared to optically thin plasmas. In non-thermal jet images, this polarization flipping appears on the photon ring\cite{tsunetoe_2024} (Fig. \ref{fig:LPflip}). Thus, we can expect to investigate the non-thermal particles through BHEX observation.

% Note: If compiling with LaTeX+dvipdf, please ensure images generated from 
% other software packages have their bounding boxes set correctly.

\subsection{Other Science Cases --- Beyond the Einstein's Theory of General Relativity}
\label{others}

In the above science cases, Einstein’s general relativity is assumed. It is also important to explore the deviation from Einstein’s theory. Recently Refs. \citenum{mizuno_2018} and \citenum{fromm_2021} studied the comparison of the shadow images of the Kerr black hole and the non-rotating dilaton black hole (i.e., beyond Einstein’s theory of general relativity) and found that these can be distinguished by using the space-VLBI. The possibility of using parameterized metrics is also explored. These are being carried out by international collaborations.

\section{Summary}

The BHEX Japan team is proposing various science cases to explore the black hole spacetime, jet formation mechanism and so on. For example, the detection of the crescent-like shadow in the brighter state in M87, the appearance of the multiple rings in the jet launching region, polarization separation and flipping on the ring, and the theory beyond Einstein’s theory. These are anticipated to be complementary to the main mission of BHEX.

\acknowledgments % equivalent to \section*{ACKNOWLEDGMENTS}       

This work was supported in part by JSPS KAKENHI grant Nos. JP23K03448, JP23H00117 (T.K.), JP21H04488 (K.O.)
This work was also supported by MEXT as “Program
for Promoting Researches on the Supercomputer
Fugaku” (Structure and Evolution of the Universe Unraveled
by Fusion of Simulation and AI; Grant Number
JPMXP1020240219 (K.O., T.K.) / Black hole accretion disks and quasi-periodic oscillations revealed by general relativistic hydrodynamics simulations and general relativistic radiation transfer calculations; Grant Number JPMXP1020240054 (K.O., T.K.), by Joint Institute for Computational Fundamental Science (JICFuS, K.O.),
and (in part) by the Multidisciplinary Cooperative Research Program in CCS, University of Tsukuba
(K.O.).
Numerical computations in this work were in part carried out on Cray XC50 at Center for Computational Astrophysics, National Astronomical Observatory of Japan.

% References
%\bibliography{SPIE_GR} % bibliography data in report.bib
%\bibliographystyle{spiebib} % makes bibtex use spiebib.bst

\end{document}